\documentclass{article}
\usepackage{emulateapj}
\usepackage{apjfonts}

\begin{document}

\submitted{To appear in The Astrophysical Journal Letters}

\title{Deep Photometry of Andromeda Reveals
Striking Similarities in the Tidal Stream and Spheroid Populations$^1$}

\author{Thomas M. Brown$^2$, 
Ed Smith$^2$,
Puragra Guhathakurta$^3$,
R. Michael Rich$^4$,
Henry C. Ferguson$^2$, 
Alvio Renzini$^5$,
Allen V. Sweigart$^6$,
\& Randy A. Kimble$^6$}

\begin{abstract}

We present a color-magnitude diagram (CMD) for a field in the giant
tidal stream of the Andromeda galaxy (M31).  These observations, taken
with the Advanced Camera for Surveys on the {\it Hubble Space
Telescope}, are 50\% complete at $V\approx 30$, reaching 1~mag below
the oldest main-sequence turnoff.  Striking similarities between the
stream and a previous spheroid CMD imply they have very similar age
and metallicity distributions, but present something of an enigma; 
we speculate on possible interpretations of
this result, but note that none are without problems. 
Distinct multiple turnoffs,
as might be expected from pulses of star formation caused by
interaction with Andromeda,  are not apparent in the stream CMD. 
Star formation in both fields lasted
about 6 billion years, building up to relatively high
metallicities and being largely complete 6 billion years ago.  
The close similarity of the spheroid and stream suggests that both 
may have derived from the same event; it would be worth
exploring to what extent stars in these structures are the remnants 
of a disk galaxy that interacted with M31, or even were disrupted from 
the M31 disk itself by the interaction.

\end{abstract}

\keywords{galaxies: evolution -- galaxies: stellar content --
galaxies: halos -- galaxies: individual (M31)}

\section{Introduction}

According to hierarchical models of galaxy formation, spheroids form in a
series of mergers between galaxies and proto-galaxies.  These models
generally predict more dwarfs than observed, leading to suggestions
that most of the proto-dwarfs in the early universe have since
dissolved into the spheroids of giant galaxies (e.g., Bullock et al.\
2000).  Until relatively recently, traces of such activity were not
obvious in the two giant galaxies of the Local Group.  However, the
discovery of the Sgr dwarf (Ibata et al.\ 1994), cannibalized by the
Milky Way, sparked renewed interest in the formation of spheroids
via the accretion of dwarfs.  Subsequently, a spectacular tidal stream
was discovered in Andromeda (Ibata et al.\ 2001), along with
a variety of substructures in the spheroid and outer disk 
(Ferguson et al.\ 2002), suggesting an active merger history. 

Further insight into the formation of galaxies can be found by studying 
their star formation histories.  Photometry
extending below the oldest main-sequence turnoff in a 
population can yield
its complete formation history, but until the advent
of the Advanced Camera for Surveys (ACS; Ford et al.\
1998) on the {\it Hubble Space Telescope (HST)}, it was not
feasible to obtain such data for populations much beyond our own Milky Way and 
its satellites.  However, Brown et al.\ (2003) used the 
ACS to obtain extremely deep photometry of a
minor-axis field in the \linebreak

{\small \noindent $^1$Based on observations made with the NASA/ESA
Hubble Space Telescope, obtained at STScI, and associated with 
proposal 10265.\\
$^2$STScI, 3700 San Martin Dr.,
Baltimore, MD 21218;  tbrown@stsci.edu, edsmith@stsci.edu, 
ferguson@stsci.edu.\\
$^3$UCO / Lick Observatory, 271 ISB,
1156 High Street, Santa Cruz, CA 95064; raja@ucolick.org.\\
$^4$Div. of Astronomy, Dpt.\ of Physics \& Astronomy, UCLA, Los
Angeles, CA 90095; rmr@astro.ucla.edu.\\
$^5$Osservatorio Astronomico, Vicolo Dell'Osservatorio 5, I-35122 Padova,
Italy; arenzini@pd.astro.it.\\
$^6$Code 667, NASA/GSFC, Greenbelt, MD 20771; \\ 
allen.v.sweigart@nasa.gov, randy.a.kimble@nasa.gov.}

\noindent
Andromeda (NGC~224; M31) spheroid; the
resulting CMD shows a dominant
intermediate-age population of 6--11~Gyr along with a significant, old,
metal-poor population.  The age distribution, high metallicity (Mould
\& Kristian 1986; Durrell, Harris, \& Pritchet 2001), and substructure
in the Andromeda spheroid all point to a violent merger history.
Given these disturbances, our use of the term ``spheroid'' in this Letter
does not imply a smooth and relaxed structure, but simply refers to
the extraplanar stars of M31.

Andromeda's tidal stream remains the most prominent merger remnant in
the Local Group, and several years of intense study have yielded
important constraints on its origin.  Ibata et al.\ (2004) found that
the stream is kinematically cold, with a velocity dispersion
($\sigma_v$) of only 11 km s$^{-1}$, and that the stream is
increasingly blue-shifted as it approaches Andromeda, implying that
the stream is falling into Andromeda from behind the galaxy.
More recently, Kalirai et al.\
(2005) found two distinct kinematic stream components -- both cold
($\sigma_v \approx 15$ km s$^{-1}$) and blue-shifted with respect to
the spheroid (Figure 1$a$).  

In all of these studies, the stream's age distribution has
remained largely unconstrained.  Ferguson et al.\ (2005) published a
stream CMD reaching a few magnitudes below the horizontal branch (HB);
they found no evidence for very young populations, but speculated that
the presence of a strong RGB bump might indicate a narrow age
dispersion.  In late 2004, we obtained deeper stream
photometry, extending 1~mag below the oldest main-sequence turnoff,
thus revealing the entire star formation history.  In this Letter, we
report on our initial analysis, which shows a remarkable similarity
between the stream population and the spheroid population of Brown et
al.\ (2003), in both the age and metallicity distributions.

\section{Observations and Data Reduction}

Using the ACS Wide Field Camera, we obtained deep optical images of a
field 1.5$^{\rm o}$ (20 kpc) from the M31 nucleus, at
$\alpha_{2000} = 00^h44^m18^s$, $\delta_{2000} = 39^{\rm
o}47^{\prime}36^{\prime\prime}$.  The field is well-placed within the
tidal stream, and is a few arcmin from the stream fields of
Ferguson et al.\ (2005), also observed with ACS.  The Keck
spectroscopy of Kalirai et al.\ (2005) is coincident with our field.
Our field includes a ``candidate''
globular cluster (Bol D242; Galleti
et al.\ 2004), intending to reach the cluster turnoff (cf.\
Brown et al.\ 2004b); unfortunately, our images show that Bol
D242 is not a cluster.  Rich et al.\ (2005, in prep.) also
obtained Keck spectra for our original spheroid field
(Brown et al.\ 2003).  Velocities in our stream (Figure
1$a$) and spheroid (Figure 1$b$) fields show that they are
kinematically distinct; approximately 77\% of the stars in our stream field
are associated with two stream components.  The stars in common between
the HST fields and the larger Keck fields follow the same velocity
distributions.

From 30 Aug to 4 Oct 2004, we obtained 14.7 hours of images in the F606W
filter (broad $V$) and 21.7 hours in the F814W filter ($I$), with every
exposure dithered to enable hot pixel removal, optimal point spread
function (PSF) sampling, smoothing of spatial variation in detector
response, and filling in the detector gap.  Our reduction process,
briefly summarized here, is very similar to that employed by Brown et
al.\ (2003), but updated to reflect the latest calibrations.  The images
were registered, rectified, rescaled to 0.03$^{\prime\prime}$
pixel$^{-1}$, and coadded using the DRIZZLE package (Fruchter \&
Hook 2002), with rejection of cosmic rays and hot pixels.  PSF-fitting
photometry, using the DAOPHOT-II software of Stetson (1987), was
corrected to agree with aperture photometry of isolated
stars, with the zeropoints calibrated at the 1\% level.  
Several improvements to the process of Brown et al. (2003) resulted in a
cleaner catalog; these include changing the detection threshold from
4$\sigma$ to 5$\sigma$ and rejecting PSF fits of poor quality
(generally due to blends of stars and superposition with background
galaxies).  After rejection of $\approx$24,000 stars, the final stream
catalog contains $\approx$100,000 stars.  Our photometry is in the
STMAG system: $m= -2.5 \times $~log$_{10} f_\lambda -21.1$.  For reference,
ABMAG~=~STMAG~$-0.169$ for
$m_{F606W}$, and ABMAG~=~STMAG~$-0.840$ mag for $m_{F814W}$. 

\section{Analysis}

The stream CMD (Figure 1$c$) looks strikingly similar to the spheroid
CMD of Brown et al.\ (2003).  For comparison, we show in Figure 1$d$
the CMD for a subset of the spheroid data that approximates the depth
and star counts in the stream data (the complete spheroid images are
$\approx$0.5~mag deeper with 40\% higher surface brightness).  The
spheroid and stream data were reduced in the same manner.  We
then shifted the spheroid 0.03~mag fainter to match the larger
distance to the stream, and 0.014~mag to the blue, to account for
small calibration errors in color (see below).
To help guide the comparison, we also show the ridge
line for NGC~104 (Brown et al.\ 2005), shifted to the distance
(770~kpc; Freedman \& Madore 1990) and reddening ($E_{B-V} =
0.08$~mag; Schlegel et al.\ 1998) of Andromeda.  The 0.03~mag
difference in $(m-M)_0$ between our stream and spheroid fields
is well within the distance uncertainties for Andromeda and 
NGC~104.  To further guide comparisons, we also show the color
distributions on the lower RGB (Figure 1$e$) and HB (Figure 1$f$), and
the luminosity distributions at the red clump (Figure 1$g$) and
subgiant branch (SGB; Figure 1$h$).

As shown in Figure 1$g$, the HB luminosities of the stream and
spheroid agree very well if the spheroid is shifted fainter by
0.03~mag (11 kpc).  The distributions are not identical, though.  The
spheroid distribution is 50\% broader, presumably due to the depth of the
flattened spheroid relative to that of the stream.  We estimate the luminosity
widths and offsets of the stream and spheroid by fitting Gaussians to these HB
distributions, with errors of $\approx$0.014~mag (5~kpc).  More
accurate distances may come from a comparison of the stream RR Lyrae
stars to those of the spheroid (Brown et al.\ 2004a), but this will be
part of a future paper.  Interpolating the results of McConnachie et
al.\ (2003) to our position would suggest that the stream is $0.14\pm
0.05$~mag ($50\pm 20$~kpc) behind the spheroid.  However, McConnachie
et al.\ (2003) estimate their stream distances by analyzing the
$I$-band luminosity functions of stars on the upper RGB; it is
plausible that they underestimate their distance errors. Support
for our relative distances comes from the data of Ferguson et al.\
(2005); we independently reduced the images of their stream, spheroid,
and ``northeast shelf'' fields.  Their stream fields are within a few
arcmin of our stream field, but their spheroid fields are
significantly further from the M31 nucleus.  The shelf is an apparent
fan of stars in the northeast quadrant of M31.  Ferguson et al.\
(2005) note that their shelf fields are 0.14~mag brighter than their
stream fields, and we concur.  Ferguson et al.\ (2005) make no
comparison between those fields and their spheroid fields, but we find
that their spheroid fields agree with our spheroid field at the
$\lesssim 0.03$~mag level, while their stream fields agree with our
stream field at the $\lesssim 0.02$~mag level.  These comparisons also
imply that McConnachie et al.\ (2003) overestimate the distance of
the stream from Andromeda, at least at a point 1.5$^{\rm o}$ from the nucleus.

Our 0.014~mag shift to the blue is well within the calibration and
reddening uncertainties; comparisons at the HB and RGB imply the shift
is not due to intrinsic age and metallicity differences.  The HB and
RGB are sensitive to metallicity and age, but not in the same sense.
As [Fe/H] increases, more stars populate the HB redward of the RR Lyrae gap,
and RGB stars become redder.  As age increases, more stars populate the HB
blueward of the RR Lyrae gap, but the RGB stars again become redder.
Our small color shift creates perfect agreement across the entire HB
and the lower RGB.  The upper RGBs are also similar, but the comparison 
is hampered by foreground contamination and the scarcity of stars.

The RGB in both fields is broad, indicating wide and similar metallicity
distributions; the RGB luminosity functions are also remarkably similar.  
The HB is dominated by a red clump, with a minority
population ($\lesssim 10$\%) of blue HB stars due to old, metal-poor
stars.  Distinct multiple turnoffs, as
might be expected from star formation pulses due to interaction of the
stream with Andromeda, are not seen.
As noted by Ferguson et al.\ (2005), the
stream RGB ``bump'' is well-defined, and extends well below the HB.
However, such a prominent bump does not necessarily indicate a narrow
age distribution, as suggested by Ferguson et al.\ (2005).
Brown et al.\ (2003) noted a similar RGB bump in their CMD of the
spheroid -- a population with broad distributions in both age and
metallicity.  This feature can be seen clearly in both of our fields
(Figures 1$c$ and 1$d$).  In globular clusters, the bump becomes
fainter (relative to the HB) as metallicity increases;
here, the bump in both the stream and spheroid slopes from a blue
point near the HB to a red point about 0.5~mag below the HB, again
indicating similarly broad distributions in metallicity for both
fields.  Such similar metallicity distributions
do not necessarily contradict the results of
Guhathakurta et al.\ (2005a), who find that the stream is slightly
more metal-rich (by $\sim$0.3 dex) than the surrounding spheroid;
their fields are further out,
where the metallicity gradient in the spheroid would enhance the
contrast with the stream.

There is a subtle difference between the stream and spheroid
luminosity functions at the SGB (Figure 1$h$).  The luminosity offset
between the HB and SGB is a standard age indicator; a 1 Gyr decrease
in age decreases this offset by $\sim$0.1~mag for ages near 10~Gyr.
When we force the HB luminosities to
agree (Figure 1$g$), the stream SGB is $\approx$0.04~mag
brighter than that of the spheroid, suggesting that the stream
population, is, on average, $\sim$300 Myr younger;  this 
would have a negligible effect on the RGB and HB.

\section{Summary and Discussion}

The stream CMD presents a challenge for the hypothesis that the
progenitor is a disrupted dwarf galaxy on an early passage close to
M31.  The dwarf apparently stopped forming stars at least 6 Gyr ago
($\lesssim$5\% of the stars in the stream field are less than 4~Gyr old).
Orbit models (e.g., Font et al.\
2005) imply that the progenitor of the stream 
only recently ($\lesssim 1$~Gyr ago) approached within 100 kpc
of M31.  Those models do not explain why star formation stopped
long before the interaction started.  

The relatively high metallicity of the stream implies that the
progenitor had a stellar mass of $\sim 10^9$ $M_\odot$ if it followed
the scaling relation for Local Group dwarfs found by Dekel \& Woo
(2003). However, according to their correlations, a dwarf galaxy with
$\sigma_v \approx 15$~km s$^{-1}$ (observed for the stream) would have
a stellar mass of only $\sim 10^6$ $M_\odot$.  Font et al.\ (2005) claim
that $\sigma_v$ can vary significantly over the orbit, and 
could be much lower than that in the progenitor;
it remains to be seen if this explanation can account for the large
mass discrepancy.  We note in this context that the only
known systems that are kinematically cold and metal rich are disks. Thus a
disk galaxy may be a more attractive candidate for the progenitor of
the stream than a pressure-supported dwarf.

Although the stream and spheroid populations have very different kinematic 
and spatial distributions, their CMDs are strikingly
similar, implying nearly identical age and metallicity distributions.
A subtle difference between the CMDs is a slightly brighter
($\approx$~0.04~mag) SGB in the stream, suggesting a slightly younger
mean age ($\sim$300 Myr).  Could the similarity of the CMDs be
explained by the stream passing through the spheroid field, as
suggested by some orbital models (e.g., Ibata et al.\ 2004)?  The
stream {\it dominates} by a 3:1 ratio in our stream CMD (Figure 1$a$;
Kalirai et al.\ 2005).  Thus the stream would
have to similarly dominate in the spheroid field, 
yet show no clear morphological signature
in the Ferguson et al.\ (2002) maps, and exhibit the kinematics shown
in Figure 1$b$ (a broad distribution at the M31 systemic velocity;
Rich et al.\ 2005, in prep.); this seems implausible given the current
orbital models.  Thus, if there is ``contamination'' from the stream 
in the spheroid field, it is likely to be much more complex and extended 
over a much wider region than that suggested by models where the 
infalling object has only completed one or two orbits.

Perhaps the stars in the spheroid came from a different dwarf galaxy
(or galaxies) that merged with M31 earlier. This hypothesis can help
explain why other locations in the spheroid (Ferguson et al.\ 2005)
show RGB and HB morphologies similar to our spheroid field (Brown et
al.\ 2003).  However, it then becomes mere coincidence that the
star-formation histories in the stream and the spheroid look nearly
identical.

Perhaps the spheroid is a disrupted disk -- either the M31 disk itself
(Brown et al.\ 2003), or the remnants of a disk galaxy that merged
with M31.  The star formation history in the spheroid might be
naturally explained if the disruption occurred 6--8 Gyr ago.  We are
still left invoking coincidence to explain the similar CMDs -- unless
the stream is also part of the disrupted disk.  It would be
interesting to explore a wider range of dynamical simulations than
those considered previously, to see whether the stream might {\it not}
be a dwarf galaxy that has encountered M31 in the last 1 Gyr, but
instead a remnant of a merged disk galaxy or a plume of stars
disrupted from the M31 disk.  It is unclear if this interpretation can
be reconciled with the stream's structure and kinematics.  Compared to
a dwarf, a disk population might better explain the extended star
formation history and high metallicity present in the stream and
spheroid.  Measurements of the formation history at
multiple locations in the stream and distorted spheroid of M31 are
undoubtedly going to be critical for sorting out what happened.
Guhathakurta et al.\ (2005b) and Irwin et al.\ (2005) find
that the minor-axis surface-brightness profile changes from a de
Vaucouleurs $r^{1/4}$ law to an $r^{-2.3}$ power law beyond $\sim$30
kpc; exploration of this region might yield important constraints on
the ``primordial'' halo unaffected by Andromeda's violent merger
history.

\acknowledgements

Support for proposal 10265 was provided by NASA through a grant from
STScI, which is operated by AURA, Inc., under NASA contract NAS
5-26555.  RMR and PG acknowledge support from the NSF.
We are grateful to P.\ Stetson for providing his DAOPHOT
code.  We thank J. Kalirai for allowing us to reproduce his velocity
distributions.

\clearpage

\epsfxsize=6.5in \epsfbox{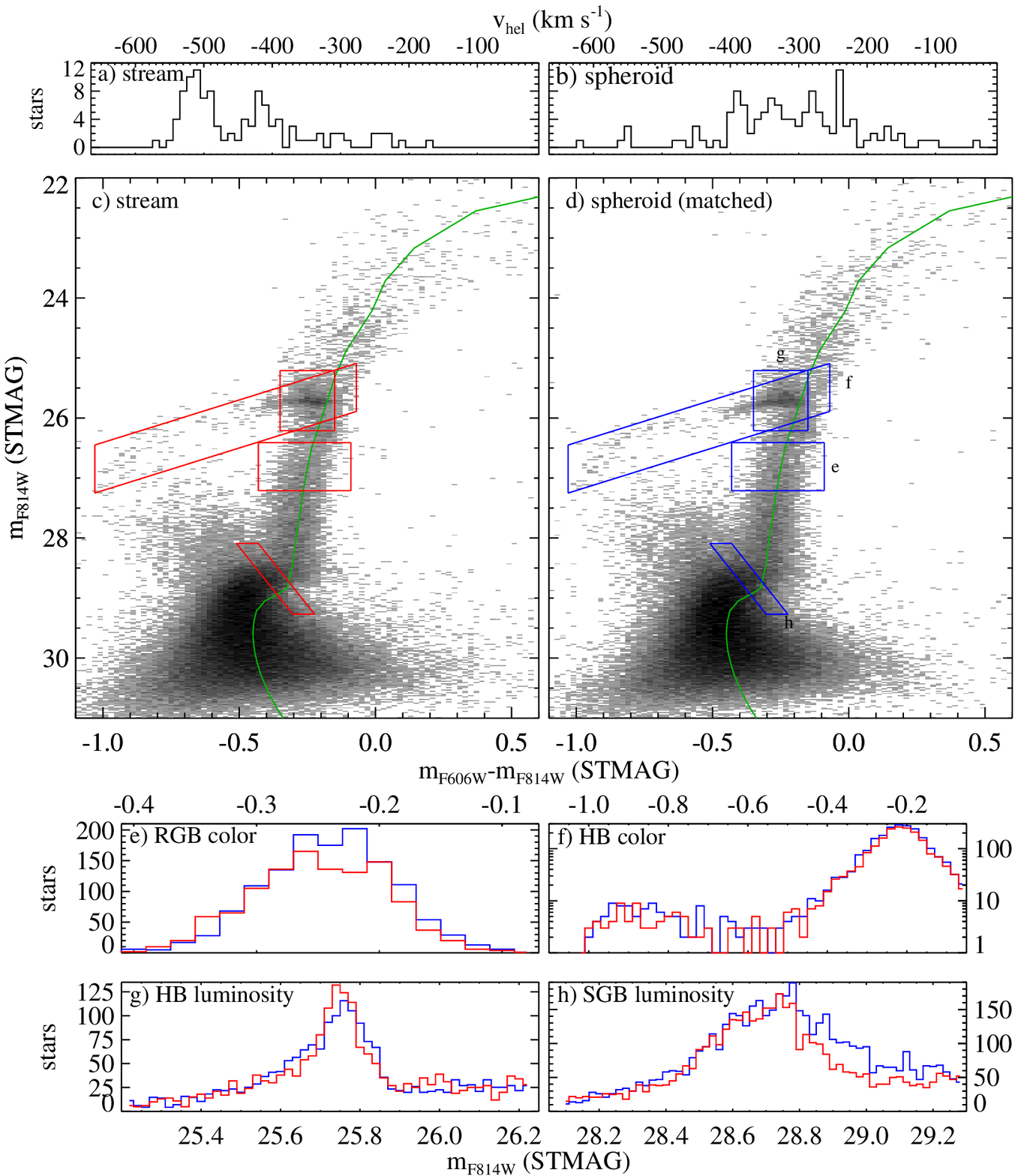} 

\vskip 0.1in

\parbox{6.5in}{\small {\sc Fig.~1--}
{\it (a)} The velocity histogram
for RGB stars at our stream position (Kalirai et al.\ 2005), showing
two cold peaks due to the stream, and a minority spheroid component
($\approx 23$\%).  {\it (b)} The velocity histogram (Rich et al.\
2005, in prep.) for RGB stars at our original spheroid position (Brown
et al.\ 2003).  Note that Ibata et al.\ (2005) recently found a cold
($\sigma_v \approx 30$ km s$^{-1}$), slowly-rotating, disk-like
structure extending to 40 kpc in the disk plane; it seems unlikely
that this structure contributes significantly here, at 50 kpc in the
disk plane, given the much broader dispersion.  {\it (c)} The CMD of
our stream field, shown as a Hess diagram with a logarithmic stretch.  
Cuts across the CMD ({\it red boxes}) are used
to highlight comparisons with the spheroid.  The ridge line for
NGC~104 (Brown et al.\ 2005) is shown for comparison.  {\it (d)} The
CMD of the spheroid, from a subset of the images chosen to match the
stream depth and a subset of the resulting catalog chosen to match the
number of stream stars.
Cuts across the CMD ({\it blue boxes}) highlight the
remarkable similarities to the stream.  Cut labels
correspond to the subsequent panels in this figure.  The spheroid has
been shifted 0.03~mag fainter and 0.014~mag to the blue (see text).
{\it (e)} Histograms for stars along the RGB color cut for the stream
({\it red}) and spheroid ({\it blue}). {\it
(f)} Histograms along the HB color cut for the stream ({\it red}) and
spheroid ({\it blue}).  {\it (g)} Histograms along the HB
luminosity cut for the stream ({\it red}) and spheroid ({\it blue}).  
{\it (h)} Histograms along the SGB luminosity cut for
the stream ({\it red}) and spheroid ({\it blue}).}

\end{document}